# High-resolution, large dynamic range fiber length measurement based on frequency-shifted asymmetrical Sagnac interferometer

**Bing Qi**

Department of Physics, University of Toronto, Toronto, Canada M5S 1A7

**Andrew Tausz and Li Qian**

Department of Electrical and Computer Engineering, University of Toronto, Toronto,

Canada M5S 3G4

**Hoi-Kwong Lo**

Department of Electrical and Computer Engineering and Department of Physics,

University of Toronto, Toronto, Canada M5S 3G4

We propose and experimentally demonstrate a single-mode fiber length and dispersion measurement system based on a novel frequency-shifted asymmetric Sagnac interferometer incorporating an acousto-optic modulator (AOM). By sweeping the driving frequency of the AOM, which is asymmetrically placed in the Sagnac loop, the optical length of the fiber can be determined by measuring the corresponding variation in the phase delay between the two counter-propagating light beams. Combined with a high-resolution data processing algorithm, this system yields a dynamic range from a few centimeters to 60km (limited by our availability of long fibers) with a resolution about 1ppm for long fibers.

*OCIS codes:* 060.2370, 060.2630, 060.5060.

Precise fiber length measurement is important in both optical communication and optical sensing. Examples include in-service fiber line identification in a complex fiber network[1], and fiber chromatic dispersion measurement[2], etc. The most common optical length measurement techniques are the optical time domain reflectometer (OTDR)[3], optical coherent domain reflectometer (OCDR)[4-5], and the optical frequency domain reflectometer (OFDR)[1,6]. These techniques are complicated to implement, and they suffer from either a small dynamic range or a low resolution. In contrast, we propose a simple approach achieving high resolution over a large dynamic range. Our approach employs a frequency shift element in an asymmetric Sagnac interferometer. The basic idea is simple: light signals of different frequencies experience different phase delays as they go through the same fiber. This phase difference, which carries information of the optical length of the fiber, can be easily measured using interference. In our setup, a polarization insensitive fiber-pigtailed AOM (Brimrose Corp.) was used to achieve the frequency shift. The acoustic wave generates a propagating diffraction grating inside the crystal. Consequently, the 1st-order diffracted light is Doppler shifted by an amount equal to the frequency of the acoustic signal $f$ [7].

Figure 1 shows our experimental setup. A 1550 nm, 2mW CW laser is used as the light source. After passing through a 2x2 symmetric fiber coupler, the laser beam is spilt into two parts equally: $S_1$, goes clockwise through the fiber loop, while $S_2$, goes through the same fiber loop counterclockwise. The frequencies of both $S_1$ and $S_2$ are up-shifted by the same amount when they come back to the fiber coupler, a stable interference signal can be observed. A computer with a Data Acquisition card is used to control the function generator (for driving the AOM) and to read the power from the photo detector. A spool of fiber with length $L_B$ (~100m) was put in the system intentionally. Also, a polarization controller was employed to improve the visibility.

Since $S_1$ and $S_2$ go through the same loop, ideally, any phase drift or polarization fluctuation will be canceled out. In practice, due to the birefringence in the loop, $S_1$ and $S_2$ may experience different phase delays, and their polarization states could also be different after they go through the loop[8]. The interference signal can be described by

$$V = (1 - m\cos\phi)/(m+1) \qquad (1)$$

Where $\phi = \phi_2 - \phi_1$ is the relative phase between $S_1$ and $S_2$, and the parameter $m \in [0,1]$ describes the visibility of the interference fringe.

Suppose the wavelength (frequency) of light before and after going through AOM are $\lambda$ $(\nu)$ and $\lambda'$ $(\nu')$ respectively. Define $\Delta\lambda = \lambda' - \lambda$ and $\Delta\nu = \nu' - \nu$. From $\lambda = C/\nu$ and $\Delta\nu = f$, we can get

$$\Delta\lambda/\lambda^2 = -f/C \qquad (2)$$

where $C$ is the speed of light in a vacuum.

The phase delays experienced by $S_1$ and $S_2$ can be expressed as

$$\phi_1 = 2\pi n L_1/\lambda + 2\pi n L/\lambda + 2\pi n L_2/\lambda' \qquad (3a)$$

$$\phi_2 = 2\pi n L_2/\lambda + 2\pi n L/\lambda' + 2\pi n L_1/\lambda' + \phi_0 \qquad (3b)$$

Where $n$ is refractive index of fiber, $L$ is the length of the test fiber, $L_1$ is the total length of the connecting fiber from the coupler to port A plus the one from port B to AOM and $L_2$ is the fiber length from AOM to coupler. Constant $\phi_0 \in [0, 2\pi)$ is introduced to take into account the phase difference caused by birefringence in the fiber loop.

Using Eqs.(1), (2) and (3), and considering $\Delta\lambda \ll \lambda$, we can get

$$V = \{1 - m\cos[2\pi n f(L+L_0)/C + \phi_0]\}/(m+1) \qquad (4)$$

Where $L_0 = L_1 - L_2$ is approximately equal to $L_B$ (100m). The interference pattern $V$ varies periodically with acoustic frequency $f$. By scanning $f$ while recording $V$, the fiber length $L$ can be determined from the "period" of $V$ with high resolution. The offset fiber $L_B$ is necessary for short fiber measurement: Without it, the required frequency scan range to complete one "period" would be too larger for the AOM.

We calculate the "period" of $V$ from the frequency difference between two minimum points $f_k$ and $f_{k+N}$ on the interference pattern. Because neither back-reflections from unwanted surfaces (which contribute to DC background in the interference pattern), nor the long term drift of optical components (such as fiber coupler) can change this "period", our system is quite robust against environment noise.

From Eq. (4), the acoustic frequency of the *k-th* minimum point in the interference pattern is

$$f_k = (2k\pi - \phi_0) \times C/[2\pi n(L+L_0)] \qquad (5)$$

So

$$f_{k+N} - f_k = NC/[n(L+L_0)] \qquad (6)$$

The fiber length $L$ can be calculated from

$$L = NC/[n(f_{k+N} - f_k)] - L_0 \qquad (7)$$

The integer $N$ in Eq. (7) can be determined by counting the number of minimums between $f_k$ and $f_{k+N}$. During the derivation of Eq.(7), the unknown constant $\phi_0$ was canceled out. Also, the parameter $m$ in Eq.(1) does not show up in Eq.(7). This means our system is insensitive to the birefringence in the fiber loop, although the use of a polarization controller can improve the visibility.

From Eq.(7), the error of the length measurement $\Delta L$ is mainly caused by $\Delta f$, which is the error in determining frequencies $f_k$ and $f_{k+N}$. Here $\Delta f$ can be separated into two parts

$$\Delta f = \Delta f_0 + \Delta f_\phi \qquad (8)$$

$\Delta f_0$ is the frequency resolution of the function generator, while $\Delta f_\phi$ is the frequency error of the data processing algorithm for fitting the minimum point from the sampling data. We assume the phase error $\delta\phi$ in finding the minimum point is independent of the fiber length. From Eq. (4)

$$\Delta f_\phi = \{C/[2\pi n(L+L_0)]\} \times \delta\phi \qquad (9)$$

By differentiating Eq. (7), and using Eq. (8) and Eq. (9), we can derive the relative resolution to be:

$$|\Delta L/L| \approx \sqrt{2} \times [f_0/(f_{k+N}-f_k)] \times [(L+L_0)/L] \times (\Delta f/f_0)$$
$$= \sqrt{2} \frac{f_0}{f_{k+N}-f_k} \times \frac{L+L_0}{L} \times \frac{\Delta f_0}{f_0} + \frac{C}{\sqrt{2}\pi n L(f_{k+N}-f_k)} \times \delta\phi \qquad (10)$$

We chose $f_k \sim$ 50MHz (the lower frequency limit of AOM) while $f_{k+N} \sim$ 56MHz (the upper frequency limit of AOM). For large $L$, the second term at the right side of Eq.(10) can be neglected, and the length resolution is limited by the frequency resolution of the function generator. For short $L$, the contribution of the phase error $\delta\phi$ cannot be neglected.

A *LabView* program was developed to scan $f$, acquire the interference fringe, search for minimum points and calculate $L$. For 60km fiber, the visibility is still about 93%. To calibrate our system, for short fibers, we used a tape measure, while for long fibers, an Agilent 86037C Chromatic Dispersion Test system, whose length resolution is 0.1%, was used. Spools of Corning SMF28 fiber (from 5m to 60km) were tested.

The length measurement results are shown in Figure 2. The relative differences between our system and the Agilent system are less than 0.1% except for a 55m fiber spool (0.18%), which we believe is due to the inaccuracy of the Agilent system for short fiber. In fact, for a 5.18m fiber (determined by tape measure), our system measured 5.20m, while the Agilent system measured 5.36m.

The resolution is defined as twice of the standard deviation. The experimental results are shown in Figure 3, which match the theoretical model very well.

The high resolution of our system suggests its potential application for chromatic dispersion measurement. In principle, by tuning the wavelength of the light source while recording the optical length, the group delay $\tau(\lambda)$ can be determined. The chromatic dispersion can be calculated from [9]

$$D(\lambda) = (\partial \tau / \partial \lambda) / L \quad (11)$$

A preliminary dispersion measurement was conducted by employing a tunable laser with a tuning range of 1480nm—1585nm. Figure 4 shows a comparison between the dispersion result obtained from our system and that from the Agilent system. The slight discrepancy may be attributed to the wavelength dependence of the components in our system, which was not calibrated.

In conclusion, we proposed and demonstrated a frequency-shifted interferometer, which can be used for high-resolution fiber length measurement. With a rather simple and robust setup, we achieved a resolution on the order of $10^{-6}$ for long fibers. We demonstrated a dynamic range of 60km, which was only limited by our availability of long fibers. By tuning the wavelength of the laser source, this system can also be used to measure chromatic dispersion.

Financial support from NSERC, CRC Program, CFI, OIT, PREA, and CIPI is gratefully acknowledged.

Fig.1. Frequency-shifted asymmetrical Sagnac interferometer for fiber length measurement.

Fig.2. Length Measurement results: "o": Calibrated by Agilent 86037C Chromatic Dispersion Test system; "*": calibrated by tape measure. The solid line is X=Y. Fiber refractive index n=1.4682 (SMF-28).

Fig.3. Resolution of our system: Circular dots indicate twice the standard deviations measured at different fiber lengths; Solid line corresponds to Eq (10) with parameters $L_0$=100m, $f_0$=53MHZ, $f_k$=50MHz, $f_{k+N}$=56MHz, $\Delta f_0 / f_0 = 5 \times 10^{-8}$ and $\delta\phi = 4 \times 10^{-4}$.

Fig.4. Chromatic dispersion measurement for a 20km SMF-28 fiber: Solid line—our system; Dashed line—Agilent 86037C Chromatic Dispersion Test system.

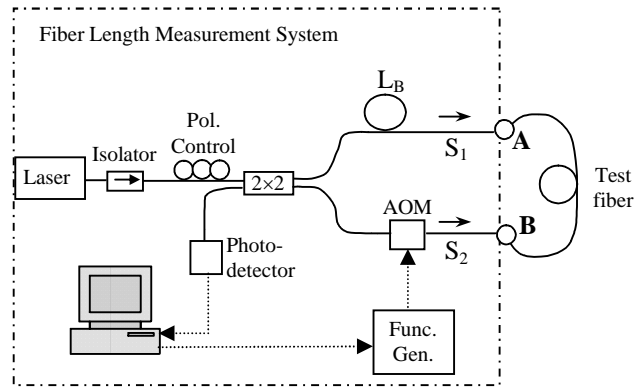

Figure.1

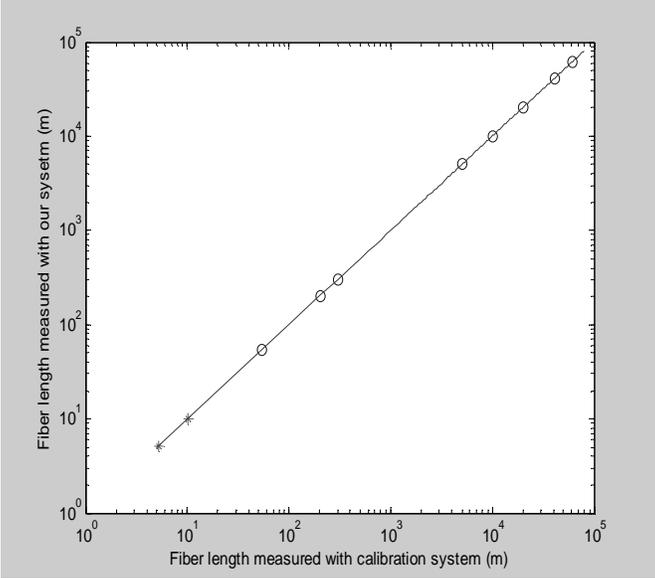

Figure.2

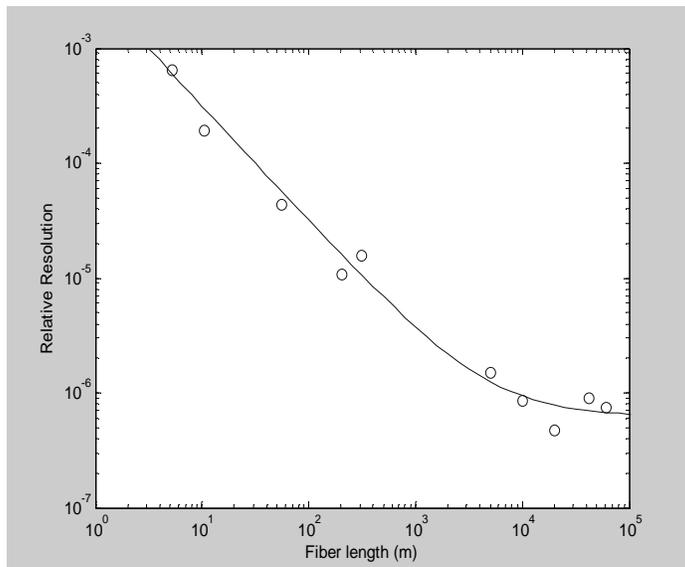

Figure.3

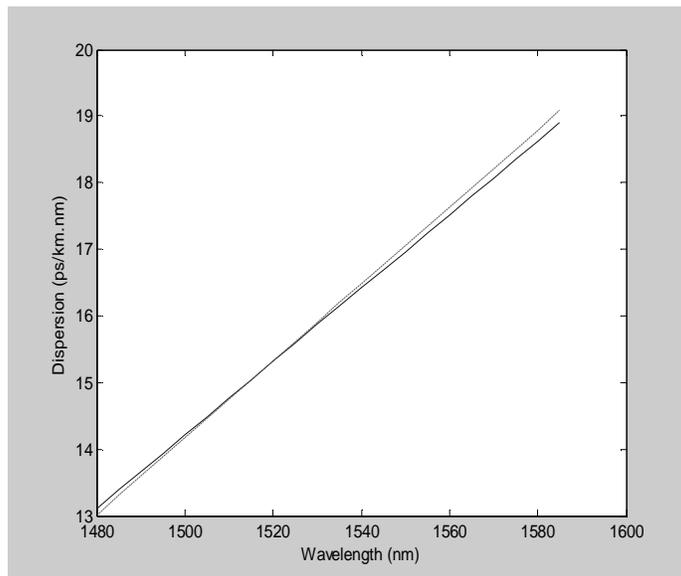

Figure.4